\documentclass[openacc]{rstransa}

\usepackage[utf8]{inputenc}
\usepackage[normalem]{ulem}

\newcommand{\beq}{\begin{equation}}
\newcommand{\eeq}{\end{equation}}
\newcommand{\bea}{\begin{eqnarray}}
\newcommand{\eea}{\end{eqnarray}}

\newcommand*{\doi}[1]{\href{http://dx.doi.org/#1}{doi: #1}}

\jname{rsta}
\Journal{Phil. Trans. R. Soc}

\begin{document}
\title{Introduction to the physics of solar eruptions and their space weather impact}
\author{Vasilis Archontis$^1$ and Loukas Vlahos$^2$}
\address{$^1$ St Andrews University, School of Mathematics and Statistics, \\St Andrews KY 16 9SS, UK\\
$^2$ Department of Physics, 
              Aristotle University, \\54124 Thessaloniki, Greece}

\subject{Astrophysics, Plasma Physics, Space Weather}


\corres{V. Archontis\\
\email{vs11@st-andrews.ac.uk}}

\begin{abstract}   
The physical processes, which drive powerful
solar eruptions, play an important role in our
understanding of the Sun-Earth connection. In this
Special Issue, we firstly discuss how magnetic fields
emerge from the solar interior to the solar surface,
to build up active regions, which commonly host
large-scale coronal disturbances, such as coronal
mass ejections (CMEs). Then, we discuss the physical
processes associated with the driving and triggering
of these eruptions, the propagation of the large-scale
magnetic disturbances through interplanetary space
and the interaction of CMEs with Earth's magnetic
field. The acceleration mechanisms for the solar
energetic particles related to explosive phenomena
(e.g. flares and/or CMEs) in the solar corona are also
discussed. The main aim of this Issue, therefore, is
to encapsulate the present state-of-the-art in research
related to the genesis of solar eruptions and their
space-weather implications.

This article is part of the theme issue ''Solar
eruptions and their space weather impact''.
\end{abstract}  
\maketitle

\section{Introduction}\label{Intro}
The ultimate driver of eruptive solar phenomena (e.g.
flares, coronal mass ejection, CMEs), which have a major
impact on space weather, is the turbulent plasma flows
in the solar convection zone that shape the structure
and dynamics of the solar atmosphere. It is important to
highlight that eruptive phenomena are dynamically very
complex and interrelated. Typically, they originate from
solar active regions (ARs) and their physical properties
and dynamics vary during the lifetime of an AR.

Space Weather predominantly encompasses the impact of solar eruptions in the heliosphere \cite{Bothemer2007}. The causal chain of physical processes that shape the structure and dynamics of eruptive events lies at the heart of the predictability of the solar magnetic activity and is vital to quantitative Space Weather forecasting. The progress needed to improve both the short and the long-term forecasting is enormous and it challenges our entire understanding of the problem \cite{Vourlidas19}. In this issue, we address four interrelated topics, namely: (i) the driving and triggering mechanisms of solar eruptions \cite{Georgoulis19, Archontis19}, 
(ii) the prediction of the geoeffective properties  of CMEs  \cite{Vourlidas19}, (iii) the Solar Energetic Particles (SEPs) \cite{Vlahos19, Anastasiadis19} and (iv) the interaction of CMEs with Earth's magnetic field \cite{Daglis19, Balasis19, Sarris19}. 

The authorship of the review articles mainly stems from the members of the Hellenic National Space Weather Research Network (HNSWR) (\url{http://proteus.space.noa.gr/~hnswrn/}) co-funded by the European Union and Greece. This Special Issue comprises the culmination of that research effort.

\section{Main topics}
In the following, we briefly describe the context of the contributed articles associated with the four main topics presented in this Special Issue.
\subsubsection{Mechanisms of Solar eruptions} 
One of the key physical process making solar magnetic activity possible is the emergence of magnetic flux from the interior of the Sun towards the surface and the outer solar atmosphere. The first review \cite{Archontis19} discusses the mechanisms of solar
eruptions originating from emerging flux regions. Observational
examples of eruptive events and numerical simulations of magnetic flux emergence present some of the most recent developments and advances related to multi-scale eruptions and ejections of hot and cool magnetized plasma (e.g. CMEs, jets) into the heliosphere. The second review \cite{Georgoulis19} focuses on the fundamental properties of CME sources and highlights a certain causal and irreversible sequence of events that occur whenever a strong (flux-massive and sheared) magnetic polarity inversion line develops in the coronal base of eruptive ARs. This irreversibility makes eruptions inevitable when certain thresholds of magnetic energy due to electric currents (that is, available for release) and magnetic helicity are crossed in these regions. This finding, originally presented in \cite{Georgoulis19}, may explain why strong polarity inversion lines cannot disappear without hosting at least one eruption.

\subsubsection{Geoffective properties of solar eruptions}
In this topic, the review by Vourlidas et al. \cite{Vourlidas19} presents some of the key CME impact parameters, which determine their geoeffectivness, such as: Time-of-Arrival (ToA), Speed of Arrival (SoA), Momentum of CME when impact the Earth's magnetic field, duration of impact,  etc. The review article focuses on how far we are form the ``acceptable'' predictions of the CME geoffective parameters, discusses the reasons that prevent us to do better at the moment and provide strategies to overcome the current open problems and challenges.

\subsubsection{Solar Energetic Particles}
Solar Energetic Particles (SEP) are an integral
part of the physical processes related to Space
Weather. The review by \cite{Vlahos19} presents the acceleration
mechanisms related to the explosive phenomena
(flares and/or CMEs) inside the solar corona. Solar eruptions cause a large scale re-configuration of the AR coronal magnetic field and host at least two well known particle acceleration mechanisms: turbulent reconnection and turbulent shocks. 

SEP events are the outcome of an impulsive and a gradual component: the former relates to the early, impulsive eruption stages dominated by magnetic reconnection and the latter is clearly the signature of the CME shock acceleration.
Impulsive and gradual events cannot easily be distinguished, as the impulsive component can be hidden under the much more intense shock-accelerated component of a gradual event.  Eruptions driven by the emerging magnetic flux are ideal sources for the impulsive injection of SEP particles in the interplanetary space \cite{Vlahos19}. The article by Anastasiadis et al. The review by \cite{Anastasiadis18} points out the current challenges to advance our physical understanding of the SEP events to the short- and long-term forecasting using empirical and physics-based methods. A list of open questions and suggestions for future work on the subject are also presented in this review.

\subsubsection{Magnetospheric, Ionospheric and Thermoshperic response to solar eruptions}
It is now accepted that Van Allen belts consist of two belts of energetic particles. The outer belt is very dynamic, composed entirely by electrons. Occasionally, a third (electron) belt is present. The outer belt electrons have been observed to reach ultra relativistic energies.  The review by \cite{Daglis19} discusses briefly the possible acceleration mechanisms for these electrons and their relation to the solar eruptions or the fast solar wind streamers. These electrons are a serious hazard to spacecrafts and have earned the dubious reputation of the ``satellite killers''. Electric fields and plasma waves, which are driven by solar eruptions, propagating through the interplanetary space and impacting the Earth are believed to be the sources of the ultra-relativistic particles. The detailed understanding of the acceleration of the ultra-relativistic electrons in the outer Van Allen belt is a key issue for Space Weather and many other astrophysical applications.

The review by \cite{Balasis19} discusses the ionospheric response to interplanetary disturbances driven by solar eruptions, using two years worth of data from the Swarm fleet of satellites surveying the Earth's top side ionosphere and measuring magnetic and electric fields. They present how to use in-situ data to study the occurrence of plasma instabilities and their impact on Space Weather.

The review article by \cite{Sarris19} is a brief overview of the key processes coupling the ionosphere and thermosphere with the magnetosphere, especially when the large interplanetary disturbances driven by solar eruptions reach and interact with Earth's magnetic field. The challenges from the state-of-the-art modeling, the missing data and our current understanding of how the impulsively driven ITM system operates is outlined.  Some of the most challenging open questions from the ITM coupling and its Space Weather impact are also presented.

\section{Summary} 

In this Special Issue we present the ongoing research, the main advances and the open questions shaping our understanding of solar eruptions and their Space Weather impact. The phenomena associated with Space Weather are a paradigm for many astrophysical plasmas. Many of the problems addressed in this issue are important space physics problems on magnetic stability, propagation of heliospheric disturbances and their interaction with planetary magnetospheres. The topics presented in this issue include contributions from the three tenets of basic research (observations, theory and modelling) and extend into the largely unexplored areas (in Heliophysics) of research-to-operations issues, including forecasting tools and assessment, forecasting and now-casting, near-real time modeling and its challenges.

\funding{Part of this study was funded by the European Union (European Social Fund) and the Greek national funds through the Operational Program ``Education and Lifelong Learning'' of the  National Strategic Reference Frame Work Research Funding Program: Thales. Investing in Knowledge society through the European Social Fund.}

\ack{We thank all the Authors for their insightful contributions, diligence and patience in helping us to prepare this Special Issue. The Guest Editors thank the Senior Publishing Editor Bailey Fallon for his sustained efforts to achieve a timely publication.}


\end{document}